\documentclass[aps,prb,twocolumn,superscriptaddress,showpacs]{revtex4}
\usepackage{graphicx}

\usepackage{dcolumn}
\usepackage{hhline}

\newcolumntype{d}[1]{D{.}{\cdot}{#1}}
\newcolumntype{.}{D{.}{.}{-1}}
\newcolumntype{,}{D{,}{,}{-1}}
\begin{document}


\title{
Power-dependent internal loss in 
Josephson bifurcation amplifiers 
}


%
\author{Michio Watanabe}
\altaffiliation{Present address: Fort Lupton Fire Protection District, 
   1121 Denver Avenue, Fort Lupton, Colorado 80621, U.S.A.}
\affiliation{RIKEN Advanced Science Institute, 
34 Miyukigaoka, Tsukuba, Ibaraki 305-8501, Japan 
}
\author{Kunihiro Inomata}
\affiliation{RIKEN Advanced Science Institute, 
34 Miyukigaoka, Tsukuba, Ibaraki 305-8501, Japan 
} 
\author{Tsuyoshi Yamamoto}
\affiliation{RIKEN Advanced Science Institute, 
34 Miyukigaoka, Tsukuba, Ibaraki 305-8501, Japan 
} 
\affiliation{NEC Nano Electronics Research Labs., 
34 Miyukigaoka, Tsukuba, Ibaraki 305-8501, Japan}
\author{Jaw-Shen Tsai}
\affiliation{RIKEN Advanced Science Institute, 
34 Miyukigaoka, Tsukuba, Ibaraki 305-8501, Japan 
} 
\affiliation{NEC Nano Electronics Research Labs., 
34 Miyukigaoka, Tsukuba, Ibaraki 305-8501, Japan}
%

\date{5 June 2009}

\begin{abstract}
We have studied nonlinear superconducting 
resonators: $\lambda/2$ 
coplanar-waveguide (CPW) resonators  
with Josephson junctions (JJs) placed in the middle  
and $\lambda/4$ CPW resonators terminated by JJs, 
which can be used 
for the qubit readout 
as ``bifurcation amplifiers." 
The nonlinearity of the resonators 
arises from the Josephson junctions, 
and because of the nonlinearity, 
the resonators with appropriate parameters 
are expected to show a hysteretic response 
to the frequency sweep,   
or ``bifurcation,"    
when they are driven with a sufficiently large power.   
We designed and fabricated resonators  
whose resonant frequencies were around 10~GHz.  
We characterized the resonators 
at low temperatures, $T<0.05$~K, 
and confirmed that they 
indeed exhibited hysteresis.  
The sizes of the hysteresis, however,   
are sometimes considerably smaller than the predictions 
based on the loaded quality factor 
in the weak drive regime. 
When the discrepancy appears, it is 
mostly explained by taking into account 
the internal loss, which 
often increases in our resonators 
with increasing drive power in the relevant power range. 
As a possible origin of the power-dependent 
loss, the quasiparticle channel of conductance 
of the JJs is discussed. 
\end{abstract}

\pacs{84.40.-x, 74.50.+r, 85.25.Cp\\
Phys. Rev. B {\bfseries 80}, 174502 (2009) 
[DOI: 10.1103/PhysRevB.80.174502]}
%


\maketitle

\section{Introduction}
\label{sec:intro}
In many experiments related to quantum information processing, 
the readout plays a critical role. 
One of the recent major steps forward 
in the experiments of 
superconducting qubits 
was related to the improvement of readout, and the quantum 
nondemolition readout was reported.\cite{Wal05}
In the readout scheme of Ref.~\onlinecite{Wal05}, 
the charge qubit was non-resonantly coupled 
to a superconducting linear resonator.   
The qubit state was detected as a shift 
in the resonant frequency of the resonator.
The measurements were done with a weak driving power 
of $n\sim1$, where $n$ is the number of measurement 
photons populated in the resonator. 

It has been suggested that employing a 
nonlinear resonator instead of linear resonator 
should relax the strong demand of 
low-noise broadband microwave measurements 
because the latching effect between the bistable states 
in a nonlinear resonator is expected to provide 
a larger but still fast enough response.\cite{Sid04}
Such scheme, or the  
``bifurcation amplifier,"
has been applied to the readout of 
charge qubits\cite{Sid06,Bou07,Met07} 
and flux qubits,\cite{Lup04,Lup06,Lee07,Lup07} where 
Josephson junctions (JJs) as nonlinear inductors 
are employed for the nonlinear resonators. 
In many cases, the nonlinear resonators with JJs 
have lumped-element capacitors, 
and their resonant frequencies are 
around 1~GHz.\cite{Lup04,Sid06,Bou07,Lee07,Lup07} 
In general, however, using a distributed element 
such as a coplanar waveguide 
(CPW)\cite{Met07,Boa07,Naa08,Ino09}  
makes it easier to increase  
the resonant frequency and the quality factor.  
A higher resonant frequency is advantageous 
for suppressing photon-number fluctuations, 
which may cause qubit dephasing 
even when the readout circuit is turned off.\cite{Ber05}   
Choosing a high quality factor and a resonant frequency 
much higher than the qubit frequency would be also helpful 
in suppressing the qubit relaxation through the 
resonator.\cite{Hou08} 

In this work, we have characterized 
at low temperatures, $T<0.05$~K, 
nonlinear CPW resonators with JJs,  
whose resonant frequencies are as high as $10-11$~GHz. 
We observed the ``bifurcation" as expected theoretically;  
the resonators showed a hysteretic response to 
the frequency sweep when they were driven with a 
sufficiently large power.  
We analyze the hysteresis in detail, 
paying close attention to the internal loss, 
which has not necessarily been examined well 
in the earlier works\cite{Lee07,Boa07} 
on superconducting nonlinear resonators 
designed for reading out flux and charge qubits.

\section{Theory}
\label{sec:Th}
\begin{figure}
\begin{center}
\includegraphics[width=0.95\columnwidth,clip]{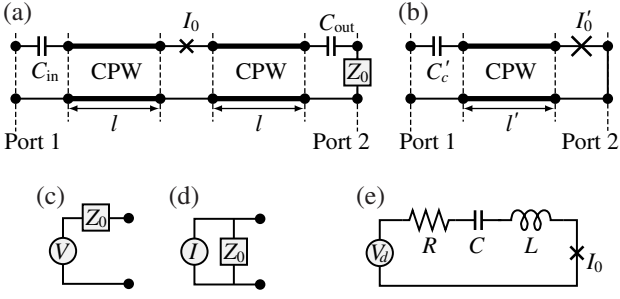}
\caption{\label{fig:circuits}
Schematic diagrams of microwave circuits. 
(a) $\lambda/2$ and (b) $\lambda/4$ coplanar-waveguide (CPW) 
resonators with Josephson junctions (JJs), where 
$I_0$ and $I_0'$ are the critical currents of the junctions, 
$l$ and $l'$ are the CPW length, and 
$C_{\rm in}$, $C_{\rm out}$, and $C_c'$ are the 
coupling capacitance to the microwave line.  
Both in (a) and (b), Port~1 is connected to a microwave source, 
which can be expressed as (c) an ideal 
voltage source and $Z_0=50$~$\Omega$ in series, 
or (d) an ideal current source and $Z_0$ in parallel. 
Port~2 is terminated by $Z_0$ in (a), 
whereas shorted in (b). 
In the vicinity of the resonance,  
(a) has been mapped to 
(e) a series circuit of $LCR$ and JJ.\cite{Boa07,Naa08}   
}
\end{center}
\end{figure}

%
The nonlinear resonators studied in this work 
are shown schematically in 
Figs.~\ref{fig:circuits}(a) and \ref{fig:circuits}(b). 
They are $\lambda/2$ and $\lambda/4$ CPW resonators, 
respectively. The crosses in the figures denote 
either single JJs or dc SQUIDs, 
the origin of the nonlinearity. We treat dc SQUIDs 
as single JJs with tunable critical currents $I_0$. 
Here, we should mention that 
superconducting transmission lines can also exhibit 
nonlinearity.\cite{Chi92,Ma97,Tho07} 
The microwave power required for this effect, 
however, is many orders of magnitude larger 
than the power range of our experiments. 
Thus, we treat superconducting CPWs 
as linear elements. 
Port~1 in Figs.~\ref{fig:circuits}(a) and 
\ref{fig:circuits}(b) is connected to a microwave source, 
whose equivalent circuits are 
Figs.~\ref{fig:circuits}(c) and \ref{fig:circuits}(d).
For uniformity, we always use Fig.~\ref{fig:circuits}(c) 
in this paper.   
The $\lambda/2$ resonator in Fig.~\ref{fig:circuits}(a) 
has been mapped\cite{Boa07,Naa08} 
to a series circuit of $LCR$ and JJ 
[see Fig.~\ref{fig:circuits}(e)] 
in the vicinity of the resonance, 
by considering the electromagnetic environment seen 
from the JJ in the linear regime, where 
the drive is so weak that the resonator does not manifest 
nonlinearity. 
The series circuit is one of the nonlinear systems 
studied by Manucharyan {\it et al.}\cite{Man07} 
They solved the equation for the Duffing oscillator   
by retaining the terms with the first harmonic only, 
then, mapped several nonlinear systems to 
the Duffing oscillator 
by comparing the differential equations of the systems  
with the Duffing equation.\cite{Man07}

%

The series resonant circuit 
in Fig.~\ref{fig:circuits}(e)  
``bifurcates" and becomes bistable in a certain 
frequency range when the drive amplitude $V_d$  
exceeds a critical value $V_c$, provided that 
the amplitude $I_c$ of the current through JJ  
at $V_d=V_c$ is sufficiently smaller than $I_0$.  
According to Ref.~\onlinecite{Man07},  
\begin{equation}
\label{eq:Vc}
V_c = \frac{8}{\,3^{3/4}\,}\left(
\frac{1}{\,Q_l\,}\frac{L}{\,L_{J0}\,}\right)^{3/2}
\omega_c \varphi_0
\end{equation}
and 
\begin{equation}
\label{eq:Ic}
I_c/I_0 = \frac{4}{\,3^{1/4}\,}\left(
\frac{1}{\,Q_l\,}\frac{L}{\,L_{J0}\,}\right)^{1/2}, 
\end{equation}
where 
\begin{equation}
\label{eq:Q}
Q_l = \left(Q_e^{-1}+Q_u^{-1}\right)^{-1}   
\end{equation}
is the loaded quality factor, $Q_e$ 
is the external quality factor, $Q_u$ 
is the unloaded quality factor,  
\begin{equation}
\label{eq:LJ0}
L_{J0} = \varphi_0/I_0     
\end{equation}
is the Josephson inductance, 
$\varphi_0 = \hbar/(2e)$, $\omega_c$ satisfies 
\begin{equation}
\label{eq:Wc}
\Omega_c \equiv 2Q_l(1-\omega_c/\omega_0) = \sqrt{3},     
\end{equation}
and $\omega_0$ is the resonant angular frequency 
in the linear regime. 
Equations~(\ref{eq:Ic}) and (\ref{eq:LJ0}) suggest 
that a large $Q_l$ and a small $I_0$ are 
favorable for the bifurcation. 
When $Q_l\gg1$, the normalized boundaries 
of the bistable region are expressed as 
\begin{equation}
\label{eq:boundaries}
\frac{V(\Omega)}{V_c} = \frac{1}{\,2\,}
\frac{\Omega^{3/2}}{\,\Omega_c^{3/2}\,}
\left[1+3\frac{\,\Omega_c^2\,}{\Omega^2} \pm 
\left( 1-\frac{\,\Omega_c^2\,}{\Omega^2} 
\right)^{3/2} \right]^{1/2}, 
\end{equation}
where $\Omega = 2Q_l(1-\omega/\omega_0)$ and 
$\omega$ is the angular frequency.\cite{Man07}

The mapping\cite{Boa07,Naa08}  
of Fig.~\ref{fig:circuits}(a) 
to Fig.~\ref{fig:circuits}(e) 
has been done for circuits with $L\gg L_{J0}$ 
by assuming that 
\begin{equation}
\label{eq:V2Vd}
V_d = Z_0 \omega_0 C_{\rm in} V,  
\end{equation}
where $Z_0=50$~$\Omega$ and 
$C_{\rm in}$ is the coupling capacitance 
to the input microwave line, 
and that 
\begin{equation}
\label{eq:Lmap}
L+L_{J0} \sim L = \pi Z_0/(2\omega_0).  
\end{equation}
Note that the mapping is based on the behavior  
in the linear regime. 
We have simulated the circuit 
in Fig.~\ref{fig:circuits}(a) taking into account 
the drive-strength dependence, 
as we will describe in the next paragraphs, 
and confirmed that 
the simulated $V_c$ and the value calculated 
from Eqs.~(\ref{eq:Vc}), (\ref{eq:V2Vd}), 
and (\ref{eq:Lmap}) 
agree reasonably well. 

The heart of our simulation is expressing 
JJ as an inductance that depends on 
the amplitude $\Delta$ of the superconducting-phase 
oscillation, 
\begin{equation}
\label{eq:Lj}
L_J (\Delta) = \frac{\Delta}{\,2J_1(\Delta)\,} \, L_{J0},   
\end{equation}
where $J$ is the Bessel function of the first kind. 
The idea is based on the following considerations: 
suppose that the superconducting phase $\delta$ of JJ 
is oscillating as  
\begin{equation}
\label{eq:delta}
\delta = \Delta\sin\omega t.   
\end{equation}
Then, the voltage $V_J$ across the junction and the current 
$I_J$ through the junction are given by 
\begin{equation}
\label{eq:Vj}
V_J / \varphi_0 = d\delta/dt = \Delta\omega\cos\omega t,    
\end{equation}
and 
\begin{eqnarray}
\label{eq:Ij}
\nonumber 
I_J / I_0 &=& \sin\delta = \sin(\Delta\sin\omega t) \\ 
&=& 2\sum_{k=0}^{\infty} J_{2k+1}(\Delta)\sin[(2k+1)\omega t].  
\end{eqnarray}
Here, let us retain the terms with the first harmonic only, 
as Manucharyan {\it et al.}\cite{Man07} did for 
solving the Duffing equation. 
Then, Eq.~(\ref{eq:Ij}) reduces to 
\begin{equation}
\label{eq:Ij_red}
I_J / I_0 = 2J_1(\Delta)\sin\omega t.  
\end{equation}
By comparing Eqs.~(\ref{eq:Vj}) and (\ref{eq:Ij_red}) 
with the equations for a usual inductor $L$, 
$V_L \propto \omega L \cos\omega t$ when $I_L \propto  
\sin\omega t$, we obtain Eq.~(\ref{eq:Lj}).   
Note that $L_J$ is {\it nonlinear} in the sense that it 
depends on $\Delta$. Its impedance $Z_J$, on the other hand, 
is written as 
\begin{equation}
\label{eq:Zj}
Z_J = j\omega L_J,
\end{equation}
as if it were a usual 
{\it linear} inductor because we retained the terms with 
the first harmonic only. 

Equations~(\ref{eq:Lj}) and (\ref{eq:Zj}) make 
the simulation easy. 
Now, for example, it is straightforward to include  
in the calculations, if necessary,  
the junction capacitance $C_J$ and the quasiparticle 
resistance $R_{\rm qp}$, both of which are parallel to $L_J$. 
The relationship between $R_{\rm qp}$ and the quality 
factors will be discussed in Sec.~\ref{sec:D}. 
We used  the transmission ($ABCD$) matrix (see, for example, 
Sec.~5.5 of Ref.~\onlinecite{Poz90}) for the simulation. 
The matrix $T_Z$ for an impedance $Z$ is 
\begin{equation}
\label{eq:Tz}
T_Z = 
\left(
   \begin{array}{cc}
   1 & Z \\
   0 & 1
   \end{array} 
\right), 
\end{equation}
and that for a section of lossless CPW 
with a length $l$ is   
\begin{equation}
\label{eq:Tcpw}
T_{\rm cpw} = 
\left(
   \begin{array}{cc}
   \cos\beta l & jZ_{\rm cpw}\sin\beta l\\
   j(Z_{\rm cpw})^{-1}\sin\beta l & \cos\beta l
   \end{array} 
\right),  
\end{equation}
where 
$j$ is the imaginary unit, 
$Z_{\rm cpw}$ is the characteristic impedance,  
$\beta = \omega/v_p$, and 
$v_p$ is the phase velocity.  
When CPW has a loss, $\beta$ 
in Eq.~(\ref{eq:Tcpw}) is replaced by 
$(\alpha+j\beta)/j$, where $\alpha$ 
characterizes the loss. 

%
\begin{figure}
\begin{center}
\includegraphics[width=0.7\columnwidth,clip]{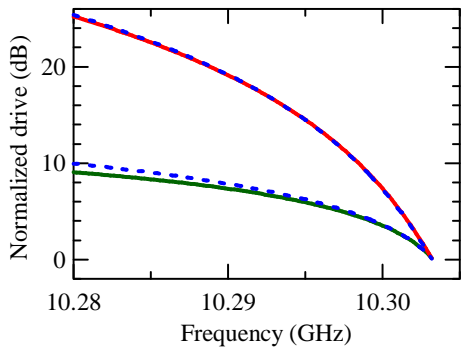}
\caption{\label{fig:sim_bound}
(Color online) 
Theoretical boundaries of the bistable region 
for a nonlinear resonator. The solid and broken  
curves are the simulation and Eq.~(\ref{eq:boundaries}), 
respectively.  
}
\end{center}
\end{figure}
%

\begin{figure}
\begin{center}
\includegraphics[width=0.95\columnwidth,clip]{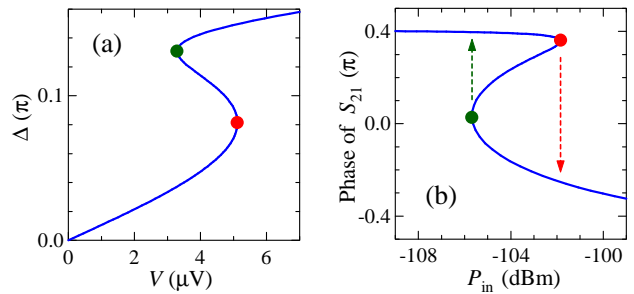}
\caption{\label{fig:sim_Pdep}
(Color online) 
Simulated drive dependence for the resonator in 
Fig.~\ref{fig:sim_bound} at $f=10.30$~GHz.  
(a) Amplitude of the superconducting-phase 
oscillation [see Eq.~(\ref{eq:delta})] vs.\ 
drive voltage amplitude. 
(b) Phase of the transmission coefficient $S_{21}$ 
vs.\ incident power to the resonator. 
The markers in both (a) and (b) correspond to 
the boundaries in Fig.~\ref{fig:sim_bound}. 
The arrows in (b) indicate 
the expected switching in actual measurements.  
}
\end{center}
\end{figure}

An example of the simulation is 
shown in Fig.~\ref{fig:sim_bound}. 
It is for a resonator of 
the Fig.~\ref{fig:circuits}(a) type. 
The parameters are 
$I_0=5$~$\mu$A, $l=2.8$~mm, and 
$C_{\rm in} = C_{\rm out} = C_c = 5$~fF. 
Both CPWs are lossless with $Z_{\rm cpw} = 50$~$\Omega$ 
and $v_p = 0.4c$, where $c$ is the speed of light. 
When we neglected $C_J$ and assumed that $R_{\rm qp}^{-1}=0$, 
the simulation yielded $f_0\equiv\omega_0/(2\pi) = 10.306$~GHz,  
$Q_l = Q_e = 3.0\times10^3$, and $V_c=2.2$~$\mu$V. 
This simulated $V_c$ and the value calculated 
from Eqs.~(\ref{eq:Vc}), (\ref{eq:V2Vd}), 
and (\ref{eq:Lmap}) agree within 2\%. 
In Fig.~\ref{fig:sim_bound}, the solid curves 
are the simulated boundaries of the bistable region,  
and compared with Eq.~(\ref{eq:boundaries}), 
the broken curves.   
The vertical axis of Fig.~\ref{fig:sim_bound} 
is the drive voltage 
amplitude $V$ normalized by $V_c$. 
In general, the solid and broken curves are similar. 
As the drive amplitude is increased, the difference 
becomes visible especially for the lower boundary, 
but this trend should be reasonable because the 
mapping of Eqs.~(\ref{eq:V2Vd}) and (\ref{eq:Lmap}) 
is based on the behavior in the linear regime   
as we mentioned earlier, 
and because at a given frequency, $\Delta$ at 
the lower boundary is actually larger than 
$\Delta$ at the upper boundary as we will see 
in Fig.~\ref{fig:sim_Pdep}(a).  

%
%

The simulated boundaries in Fig.~\ref{fig:sim_bound} 
were obtained by calculating $\Delta$ vs.\ $V$ at  
different frequencies. 
Such calculation for $f = 10.30$~GHz is shown 
in Fig.~\ref{fig:sim_Pdep}(a); 
the marked local maximum and minimum of $V$     
are the boundaries.  
In experiments, we measure instead of $\Delta$, 
the transmission coefficient $S_{21}$ of 
$\lambda/2$ resonators 
and the reflection coefficient $\Gamma$ of 
$\lambda/4$ resonators. 
These quantities are also simulated by the same method. 
The phase of $S_{21}$ is computed 
in Fig.~\ref{fig:sim_Pdep}(b) as a function of 
the incident power $P_{\rm in}$, 
which is more relevant than $V$ 
from the experimental point of view. 
The power is given by 
\begin{equation}
\label{eq:Pin}
P_{\rm in}=V_{\rm in}^2/(2Z_0),
\end{equation}
where 
\begin{equation}
\label{eq:Vin}
V_{\rm in} = V/2 
\end{equation}
is the amplitude  
of the incident voltage.  
Because the section between the markers 
is unstable, the resonator response 
in actual measurements is expected to be 
hysteretic, that is, switching indicated by the arrows 
in Fig.~\ref{fig:sim_Pdep}(b) should be observed.  

Regarding $\lambda/4$ resonators, 
when we chose 
in Fig.~\ref{fig:circuits}(b), 
$I_0'=2I_0$, $C_c'=C_c$, $l'=l$, and the 
same unit-length CPW properties,   
the simulation yielded essentially the same 
$\omega_0,$ $Q_l,$ $V_c,$ etc.\
as those of the $\lambda/2$ resonator 
in Figs.~\ref{fig:sim_bound} and \ref{fig:sim_Pdep}. 

When we compare the theoretical predictions 
in this section with 
experimental results, we should note that in 
experiments, we cannot make the drive amplitude 
arbitrarily large in order to see the bistability 
arising from the nonlinear inductance of JJs because 
for example, $I_J$ has to be smaller than $I_0$. 
Let us also recall that both Eq.~(\ref{eq:boundaries}) 
and the simulation are based on the approximation of 
retaining the terms with the first harmonic only;    
the accuracy of the approximation becomes worse as 
the drive amplitude increases.

\section{Experiment}
\label{sec:Ex}
We studied two series of CPW resonators listed in 
Table~\ref{tab}.  
The fabrication of Series~A, 
which has single JJs, 
was almost the same as that of the sample 
in Ref.~\onlinecite{Ino09};   
a Si wafer covered by a layer of 
thermally oxidized SiO$_2$ was used,   
Nb interdigital coupling capacitors and Nb CPWs were 
patterned by photolithography and reactive ion etching, 
and Al JJs by electron-beam lithography and shadow evaporation.
In order to realize a superconducting contact 
between Nb and Al, the surface of Nb was cleaned 
by Ar$^+$ milling before the Al evaporation.
In Series~B with dc SQUIDs, on the other hand, 
all electrodes including those of SQUIDs are Nb, 
and everything was fabricated 
by the photolithographic technology. 
Thus, between the two series of resonators, 
there are a couple of differences regarding the CPW 
that we should note:   
the thickness of Nb film and the quality of SiO$_2$.  

%
\begin{table}
\caption
{\label{tab}
List of resonators. A1--A4 have single Josephson 
junctions, whereas B1--B2 have dc SQUIDs. 
$f_0$ is the resonant frequency in the linear regime,   
$Q_l$ is the loaded quality factor, 
and $P_c$ is the critical incident power 
for the bifurcation.
All values are for 
the zero magnetic field, 
and $Q_l$ was calculated in the linear regime.   
}
\begin{ruledtabular}
\begin{tabular}{ccccc}
Reso-	& &$f_0$&$Q_l$&$P_c$\\
nator & Type & (GHz)&($\times$10$^3$)&(dBm)\\
 \hline
A1& $\lambda/2$ & \phantom{1}9.86 & 6\phantom{.0} 
	& \phantom{1}$-$97 $\pm$ 2  \\
A2& $\lambda/4$ & \phantom{1}9.70 & 3\phantom{.0} 
	& $-$100 $\pm$ 1  \\
A3& $\lambda/2$ & \phantom{1}9.98 & 1.1           
	& $-$101 $\pm$ 2  \\
A4& $\lambda/4$ & \phantom{1}9.83 & 1.6           
	& $-$103 $\pm$ 3  \\
B1& $\lambda/2$ &          11.30 & 1.0 
	& \phantom{1}$-$85 $\pm$ 1  \\
B2& $\lambda/4$ &          11.12 & 1.3 
	& \phantom{1}$-$82 $\pm$ 1  \\
\end{tabular}
\end{ruledtabular}
\end{table}
%

The Nb film is much thinner in Series~A 
(0.05~$\mu$m) than in Series~B (0.4~$\mu$m). 
As a result, highly likely due to the kinetic 
inductance,\cite{Mes69} we saw a noticeable difference 
in $v_p$. In Series~A, Nb is deposited directly 
on the thermally oxidized SiO$_2$ of the Si wafer, 
whereas in Series~B, there is a layer of sputtered 
SiO$_2$ between Nb and thermally oxidized SiO$_2$. 
It seems that this sputtered SiO$_2$ decreases $Q_u$.   
According to Ref.~\onlinecite{OCo08}, 
the loss tangent of SiO$_2$ grown by 
plasma-enhanced chemical vapor deposition (PECVD) 
is much larger than that of thermally oxidized 
SiO$_2.$ Thus, it is highly likely that 
sputtered SiO$_2$ is also much more lossy 
than thermally oxidized SiO$_2.$ 
We will come back to this point in Sec.~\ref{subsec:SQ}.

Our $\lambda/2$ resonators (A1, A3, and B1) 
are designed to be symmetric, 
that is, in Fig.~\ref{fig:circuits}(a), 
$C_{\rm in} = C_{\rm out} = C_c$ 
and the two CPWs are the same.
For all resonators, 
we intended to obtain $f_0\sim10$~GHz 
and $Q_l$ on the order of $10^3$ with $l,$ $l'\sim 3$~mm,  
by aiming at   
$I_0$, $I_0' \sim 3-30$~$\mu$A, 
$Z_{\rm cpw} \sim 50$~$\Omega$,  
$v_p \sim 0.4c$, and   
$C_c,$ $C_c'\sim 2-10$~fF.

The resonators were characterized in 
a $^3$He-$^4$He dilution refrigerator at 
the base temperature, $T<0.05$~K. 
For the characterization, we used  
a vector network analyzer choosing an intermediate 
frequency (IF) between 100~Hz and 40~kHz.
The transmission coefficient 
$S_{21}$ of $\lambda/2$ resonators 
were measured by connecting Ports~1 and 2 
in Fig.~\ref{fig:circuits}(a) to 
the network analyzer. 
In the input microwave line connected to Port~1,  
20-dB attenuators were inserted 
at $T=1$~K and at the base temperature. 
The output line had an isolator at $T=1$~K, 
and at $T=4$~K, an cryogenic amplifier, 
whose gain at 10--11~GHz was 34--38~dB 
or 40~dB. 
For measuring the reflection coefficient 
$\Gamma$ of $\lambda/4$ resonators, 
we employed a directional coupler and extra isolators 
at the base temperature as in  
Fig.~1(a) of Ref.~\onlinecite{Ino09}. 
Magnetic fields were applied in the direction 
perpendicular to the substrate by means of a 
superconducting solenoid. 

The gains of amplifiers and the insertion losses 
of attenuators and cables had been characterized 
at relevant frequencies by independent measurements. 
The uncertainties of  
$P_{\rm in}$, $|S_{21}|$, and 
$|\Gamma|$ in this paper are less than 1--2~dB.

\section{Results}
\label{sec:R}
%
Basic parameters of the resonators are 
listed in Table~\ref{tab}. 
We obtained $f_0\sim10$~GHz 
and $Q_l$ on the order of $10^3$ as we designed.  
Hystereses were observed in all resonators; 
the critical incident power $P_c$ 
in Table~\ref{tab} was determined experimentally. 
The values of $P_c$ $[=(V_c/2)^2/(2Z_0)]$ 
for Series~A correspond to $I_0$, $I_0'=5-21$~$\mu$A, 
or the critical current density $J_c = 1-2$~kA/cm$^2$,  
within the theory\cite{Man07} and the 
mapping\cite{Boa07,Naa08} in Sec.~\ref{sec:Th}.     
The estimated critical current 
is on the right order of magnitude, and consistent 
with the dc measurements on test junctions fabricated 
on the same wafer, $J_c\sim1$~kA/cm$^2$.  
We also estimated the critical current 
by the simulation in Sec.~\ref{sec:Th} 
taking into account the junction capacitance, 
and confirmed that the estimates 
did not change significantly. 
The situation is similar in Series~B.  
The estimated $J_c = 0.5-0.6$~kA/cm$^2$ 
is on the right order of magnitude, 
and consistent with the parameters of 
the fabrication process, which targeted at 
$J_c = 0.4$~kA/cm$^2$. 
Below, we focus on Resonators~A1 and B1 
because within each series, 
the results were qualitatively similar.

\subsection{
Resonator with a single Josephson junction
}
\label{subsec:sJJ}
%

\begin{figure}
\begin{center}
\includegraphics[width=0.95\columnwidth,clip]{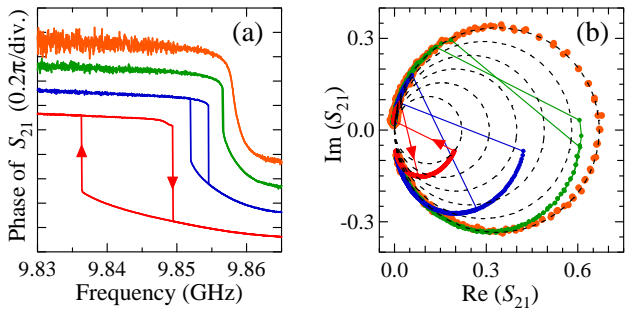}
\caption{\label{fig:A1Pdep}
(Color online) 
Dependence on the incident power $P_{\rm in}$ of  
Resonator~A1. In both (a) and (b), from right to left, 
$P_{\rm in}=-101$, $-93$, $-85$, and $-75$~dBm.   
(a) Phase of the transmission coefficient $S_{21}$ 
vs.\ frequency. The frequency is swept up and down 
for all curves. For the 
$P_{\rm in}=-75$~dBm curve, the sweep directions 
are indicated by the arrows. 
The origin of the vertical axis is 
offset for each curve for clarity. 
(b) Imaginary part of $S_{21}$ vs.\ 
real part of $S_{21}$. 
The diameters $d$ of the dotted circles range from  
0.22 to 0.67 in steps of 0.09. 
The centers are located at ($d/2,$ 0). 
}
\end{center}
\end{figure}

\begin{figure}
\begin{center}
\includegraphics[width=0.75\columnwidth,clip]{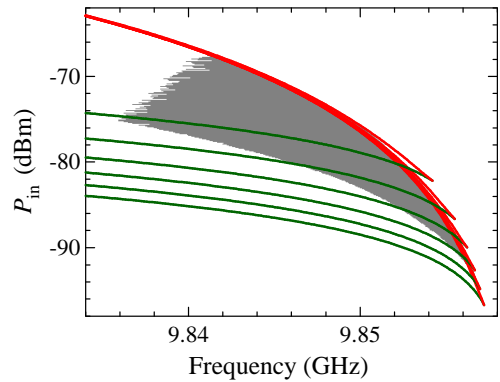}
\caption{\label{fig:A1bound}
(Color online) 
Bistable region (grey) of Resonator~A1. 
The pairs of curves are Eq.~(\ref{eq:boundaries})  
with different loaded quality factors $Q_l.$ 
As the power $P_{\rm in}$ is increased, 
the theoretical curves with a smaller $Q_l$ 
reproduce the experimental bistable region. 
}
\end{center}
\end{figure}

The $P_{\rm in}$ dependence of Resonator~A1 is 
shown in Fig.~\ref{fig:A1Pdep}. At different 
values of $P_{\rm in}$, the frequency $f$ was swept 
up and down while $S_{21}$ was recorded. 
At $P_{\rm in}=-101$~dBm $<P_c,$ that is, 
in the linear regime, the phase of 
$S_{21}$ vs.\ $f$ shows a usual rotation 
[the top right curve in Fig.~\ref{fig:A1Pdep}(a)],  
and the polar plot of Im$(S_{21})$ vs.\ 
Re$(S_{21})$ falls on the largest dotted circle 
in Fig.~\ref{fig:A1Pdep}(b). The center of the 
circle is at ($d/2$, 0), where $d$ $(=0.67)$ is 
the diameter. Note that $d$ is equal to the maximum 
$|S_{21}|.$ It is also related to 
the quality factors by 
\begin{equation}
\label{eq:d}
Q_e/Q_u = d^{-1} - 1. 
\end{equation}
Thus, $Q_e/Q_u=0.5$ for Resonator~A1 in the 
linear regime. 
The other curves in Fig.~\ref{fig:A1Pdep} are for 
$P_{\rm in}>P_c,$ and hysteresis is seen as 
the theory predicts. 
An interesting feature in Fig.~\ref{fig:A1Pdep}(b) 
is that the three curves for $P_{\rm in}>P_c$ 
do not always stay on the largest circle. 
They move towards smaller circles 
as $|S_{21}|$ increases, and the trend is stronger 
at larger $P_{\rm in}.$ 
This $P_{\rm in}$ dependent quality factors are a key 
to understand Fig.~\ref{fig:A1bound}, where 
the measured bistable region in grey is compared 
with Eq.~(\ref{eq:boundaries}).    

%
%

The grey area in Fig.~\ref{fig:A1bound} 
is determined from many two-directional 
frequency sweeps like Fig.~\ref{fig:A1Pdep}(a). 
If the phase difference of $S_{21}$ between 
up and down frequency sweeps at ($f,$ $P_{\rm in}$) 
was more than $0.1\pi,$ ($f,$ $P_{\rm in}$) 
in Fig.~\ref{fig:A1bound} is grey. 
The pairs of curves in Fig.~\ref{fig:A1bound} 
are Eq.~(\ref{eq:boundaries})  
with different values of $Q_l.$ 
The pair with the largest $\omega_c$ 
is for $Q_l = Q_{l0}$ in the linear regime 
and associated with 
the largest circle in Fig.~\ref{fig:A1Pdep}(b). 
The other pairs in Fig.~\ref{fig:A1bound} correspond  
to the smaller circles in Fig.~\ref{fig:A1Pdep}(b) 
when we neglect the $P_{\rm in}$ dependence of $Q_e$, 
which is expected to be much smaller than 
that of $d$ in Resonator~A1. 
The ratio of $V_c(Q_l)/V_c(Q_{l0})$ in Fig.~\ref{fig:A1bound} 
is consistent with Eqs.~(\ref{eq:Vc}) and (\ref{eq:Wc}), 
that is, the ratio is equal to 
$(Q_l/Q_{l0})^{-3/2}(1-\sqrt{3}/Q_l)/(1-\sqrt{3}/Q_{l0}).$ 
We treated $V_c(Q_{l0})$ as an adjustable parameter, 
and chose its value so that the experimental bistable 
region stayed between the theoretical boundaries.

The theoretical size of the bistable region 
in Fig.~\ref{fig:A1bound} shrinks rapidly 
with decreasing $Q_l$, 
and as we have seen, Fig.~\ref{fig:A1Pdep}(b) 
suggests that $Q_l$ of the resonator decreases 
with increasing $P_{\rm in}$ especially around 
the lower boundary, where $|S_{21}|$ takes the 
maximum value.  
Thus, $P_{\rm in}$ dependent $Q_l$ 
should mostly explain the fact that 
the experimental bistable region is 
considerably smaller than 
the theoretical prediction for $Q_l=Q_{l0}.$   
In general, the size of experimentally determined 
bistable region may depend on the sweep speed 
due to finite lifetimes of the bistable states. 
This effect, however, seems to be not significant 
in Fig.~\ref{fig:A1bound} judging from the experiment 
described in the following paragraphs.

%
%

We swept $P_{\rm in}$ instead of $f$ 
in two directions with ${\rm IF}=40$~kHz, 
which is much faster than ${\rm IF}=200$~Hz 
in Fig.~\ref{fig:A1bound}. 
We chose $f=9.848$~GHz and 9.853~GHz, 
where the widths of the bistable region 
in Fig.~\ref{fig:A1bound} are 8.1~dB and 5.2~dB, 
respectively.   
At both frequencies, the values of $P_{\rm in}$ 
at which the phase of $S_{21}$ switched 
[see also the arrows in Fig.~\ref{fig:sim_Pdep}(b)] 
agreed with Fig.~\ref{fig:A1bound} within 0.1~dB, 
despite the large difference in the sweep speed.  
This is the main reason why we presume   
that the dependence on the sweep speed is insignificant 
in Fig.~\ref{fig:A1bound}.

%
\begin{figure}
\begin{center}
\includegraphics[width=0.95\columnwidth,clip]{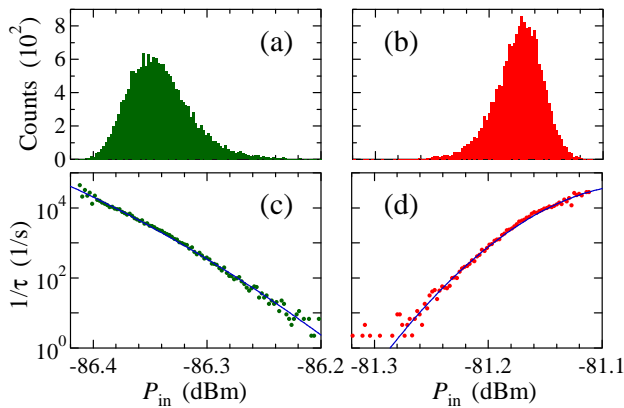}
\caption{\label{fig:histogram}
(Color online) 
Switching between the bistable states at 
$f=9.853$~GHz in Resonator~A1. 
(a) and (b) Histograms of the switching events. 
The horizontal axis is the incident power $P_{\rm in}$.  
(c) and (d) Inverse of $\tau$ vs.\ $P_{\rm in}$, 
where $\tau$ is the lifetime of the bistable state 
calculated from (a) and (b), respectively. 
The curves are the least-squares fittings of 
Eqs.~(\ref{eq:tau}) and (\ref{eq:Pb}). 
}
\end{center}
\end{figure}
%

At $f=9.853$~GHz, the measurements were 
repeated $1.8\times10^4$ times, and 
the histograms of the switching events 
are shown in Figs.~\ref{fig:histogram}(a) 
and \ref{fig:histogram}(b). 
The distribution widths of the switching events 
are much smaller than 
the width of the bistable region.   
The histograms can be converted to the lifetimes $\tau$ 
of the bistable states 
by assuming that the probability  
of remaining in the same state decays as 
$\exp(-\Delta t/\tau)$, where $\Delta t$ is the time and 
in our case, equal to the inverse of IF. 
This type of conversion has been done many times 
for the conventional switching from the zero-voltage state 
to the voltage state in JJs, and the details of the 
conversion are found, for example, 
in Ref.~\onlinecite{Ful74}. 
The lifetimes obtained from Figs.~\ref{fig:histogram}(a) 
and \ref{fig:histogram}(b) are shown in 
Figs.~\ref{fig:histogram}(c) and \ref{fig:histogram}(d), 
respectively. Below, we analyze $\tau$ within 
a simple model.

Let us suppose that  
\begin{equation}
\label{eq:tau}
\tau^{-1} \propto\exp(-E/k_BT),  
\end{equation}
and that as in Ref.~\onlinecite{Kau88}, 
$E$ is associated with the noise.  
When we further assume for simplicity that 
$E$ is proportional to the minimum $V_N^2,$
where $V_N$ is the noise voltage satisfying 
$V_N+Ve^{j\theta_N}=V_b$,  
$V$ is the source voltage, 
$\theta_N$ is the phase noise, and 
$V_b$ corresponds to 
one of the theoretical boundaries of 
the bistable region [see the markers 
in Fig.~\ref{fig:sim_Pdep}(a)],   
the power dependence of $E$ is written as 
\begin{equation}
\label{eq:Pb}
E \propto \left(\sqrt{P_{\rm in}} 
- \sqrt{P_b}\right)^2      
\end{equation}
because $V$ and $V_b$ are in phase when 
$V_N^2$ takes the minimum value. 
Intuitively speaking, 
the switching between the bistable states would occur  
at $P_b$ if an experiment with no noise were possible. 
The last assumption is actually consistent with 
the approach based on the Poincar\'{e} section 
in Fig.~2 of Ref.~\onlinecite{Sid04} because the height of 
the saddle point in the Poincar\'{e} section is 
proportional to the minimum $|V_N|$. 

The curves in Figs.~\ref{fig:histogram}(c) 
and \ref{fig:histogram}(d) are the least-squares 
fittings of Eqs.~(\ref{eq:tau}) and (\ref{eq:Pb}). 
The fittings yielded $P_b = P_l - 0.5$~dB 
and $P_h + 0.1$~dB. 
Because $P_b$ extends the bistable region 
by 0.6~dB only in total, 
the above analysis of $\tau$ at $f=9.853$~GHz 
supports our presumption that the dependence 
on the sweep speed 
is insignificant in Fig.~\ref{fig:A1bound}.

%
\subsection{
Resonator with a dc SQUID 
}
\label{subsec:SQ}
%

%
\begin{figure}
\begin{center}
\includegraphics[width=0.9\columnwidth,clip]{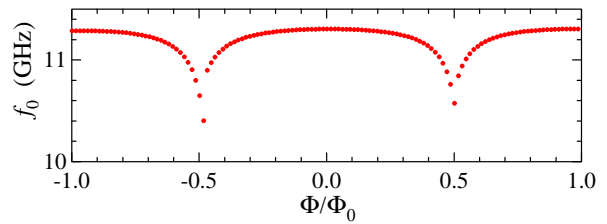}
\caption{\label{fig:f0}
(Color online) 
Resonant frequency $f_0$ in the linear regime vs.\ 
external dc magnetic 
flux $\Phi$ normalized by the superconducting 
flux quantum $\Phi_0=h/(2e)$ 
for Resonator~B1. 
The direction of the flux sweep was 
from left to right. 
}
\end{center}
\end{figure}
%

The SQUID modulation of Resonator~B1 
in the linear regime measured at 
$P_{\rm in} = -113$~dBm 
is shown in Fig.~\ref{fig:f0}. 
The external dc magnetic flux $\Phi$
indeed changes $I_0$, and thereby, 
$L_{J0}$ and $f_0$, periodically. 
Around $\Phi/\Phi_0=\pm0.5,$ the response 
was a little hysteretic and $f_0$ depended 
on the direction of the flux sweep. 
In Fig.~\ref{fig:f0}, the flux was swept in 
one direction only, and that is why 
there is small asymmetry at $\Phi/\Phi_0\sim\pm0.5.$ 
From the amplitude of the $f_0$ modulation, 
we estimate based on the simulation in Sec.~\ref{sec:Th} 
that at $\Phi/\Phi_0\sim\pm0.5,$ the critical current 
of the SQUID becomes about 15\% of its maximum value, 
and thus, the SQUID has a normalized loop inductance 
of $L_{\rm SQ}I_{00}/\Phi_0\sim0.1,$ where $I_{00}$ 
is the critical current per junction at $\Phi=0.$ 
The above value of $L_{\rm SQ}I_{00}/\Phi_0$ is 
consistent with the design.   

\begin{figure}
\begin{center}
\includegraphics[width=0.95\columnwidth,clip]{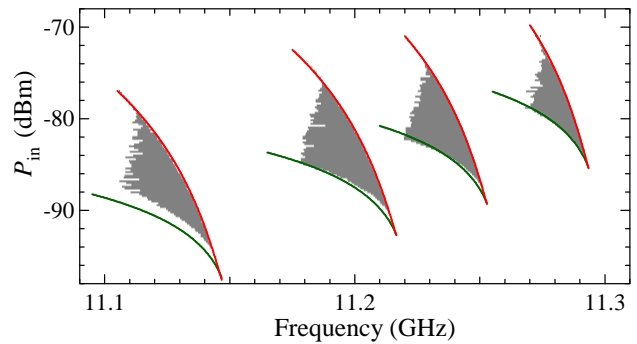}
\caption{\label{fig:B1bound}
(Color online) 
Bistable regions (grey) of Resonator~B1 
when the resonant frequencies $f_0$ in the 
linear regime were tuned to 
11.16, 11.23, 11.26, and 11.30~GHz  
by dc magnetic flux.  
The pairs of curves are Eq.~(\ref{eq:boundaries}), 
where the loaded quality factors from left 
to right are 
$Q_l/10^3 =1.2$, 1.2, 1.2, and 1.3. 
}
\end{center}
\end{figure}

For different values of $f_0,$ we measured 
the bistable regions. Some of the results 
are shown in Fig.~\ref{fig:B1bound}. 
The rightmost data set 
was obtained when $f_0$ was close to its maximum value.  
Regarding quality factors, we found two major 
differences compared to Resonator~A1.  
One is that $Q_e/Q_u$ $(\sim 5)$ in the linear 
regime is much larger. This should be due to 
the layer of sputtered SiO$_2$ that exists only 
in Series~B as we mentioned 
in Sec.~\ref{sec:Ex}.  
The other difference is that judging from 
polar plots (data not shown) like Fig.~\ref{fig:A1Pdep}(b), 
$Q_l$ does not change very much when $P_{\rm in}$ 
is increased from $P_c$.  
Actually, we estimate at $P_{\rm in} \sim P_c$, 
a larger value of $Q_l=1.3\times10^3$ than 
that in the linear regime. 
The two curves for the rightmost data set 
in Fig.~\ref{fig:B1bound} are Eq.~(\ref{eq:boundaries}) 
calculated with $Q_l=1.3\times10^3$. 
They reproduce the experimental boundaries of 
the bistable region.

%
%

The other data sets 
in Fig.~\ref{fig:B1bound} 
for smaller values of $f_0$ 
are also compared with Eq.~(\ref{eq:boundaries}), 
where we employed for $Q_l$, the estimated values  
at $P_{\rm in} \sim P_c$ again. 
The dependence of $P_c$ on $f_0$ 
qualitatively agrees with Eq.~(\ref{eq:Vc}) 
because a smaller $f_0$ means a 
larger $L_{J0}$. 
As $f_0$ is decreased, however, 
it becomes easier to see 
some discrepancies between 
the experimental boundaries and 
Eq.~(\ref{eq:boundaries}). 
In the leftmost data set, 
for example, 
polar plots (data not shown) like 
Fig.~\ref{fig:A1Pdep}(b) suggest 
a moderate decrease of 
$Q_l$ with increasing $P_{\rm in}$ 
above $P_{\rm in} \sim P_c$. 
Qualitatively speaking, the behavior 
of $Q_l$ is becoming similar to that 
in Resonator~A1.

\begin{figure}
\begin{center}
\includegraphics[width=0.65\columnwidth,clip]{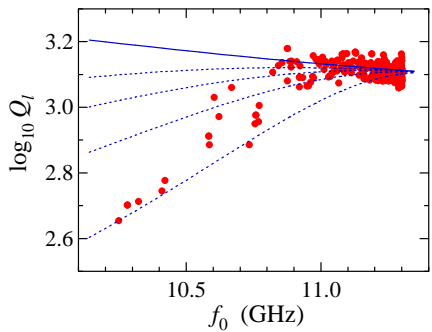}
\caption{\label{fig:B1Ql}
(Color online) 
Logarithm of the loaded quality factor $Q_l$ 
as a function of the resonant frequency $f_0$ 
in the linear regime for Resonator~B1. 
The solid and dotted curves are the theoretical predictions 
based on the simulation in Sec.~\ref{sec:Th}. 
The solid curve is for $R_{\rm qp}^{-1}=0$, that is, 
the quasiparticle resistance $R_{\rm qp}$ of the SQUID 
is not taken into account, 
whereas the dotted curves from top to bottom are for 
$R_{\rm qp}=20$~k$\Omega$, 10~k$\Omega$, 5~k$\Omega$, 
and 2~k$\Omega$, respectively.  
}
\end{center}
\end{figure}

When $f_0$ is decreased further down to $f_0<11$~GHz,  
we did not find the bistable region at all,  
which is a bit surprising because according 
to Eq.~(\ref{eq:Ic}), a larger $L_{J0}$ 
is more advantageous for observing the bistability, 
in principle. 
We should not forget, however, that 
$I_c/I_0$ in Eq.~(\ref{eq:Ic}) 
depends on $Q_l$ as well, and 
in fact, $Q_l$ at $f_0<11$~GHz decreases 
with decreasing $f_0$ 
rather sharply even in the linear regime 
as shown in Fig.~\ref{fig:B1Ql}, where 
the vertical axis is the logarithm of $Q_l,$ 
and $Q_l$ was calculated from the data 
at $P_{\rm in} = -113$~dBm.  
We will discuss the origin of the change 
in $Q_l$ later in the next section. 
The lifetimes $\tau$ of the bistable states 
may be also related 
to the disappearance of the bistable region 
at $f_0<11$~GHz.   
As $f_0$ is decreased, the relevant power 
range becomes smaller. Within the simple 
model of Eqs.~(\ref{eq:tau}) and (\ref{eq:Pb}), 
smaller $P_{\rm in}$ and $P_b$ make $E$ smaller, 
and thereby, $\tau$ shorter. 
Thus, we cannot exclude the possibility that 
our measurements were not fast enough 
for observing the bistability at $f_0<11$~GHz.

%
%

\section{Discussion}
\label{sec:D}
%

In the preceding section, we have seen 
that how the quality factor depends on 
$P_{\rm in}$ or $\Phi$ is relevant to the 
experimental bistable region. 
In this section, let us discuss the origin of 
the dependence. 

Experimental determination of all the quality 
factors in Eq.~(\ref{eq:Q}) is straightforward 
in the linear regime. 
For $\lambda/2$ resonators, 
$Q_l$ is obtained from $|S_{21}|$ vs.\ $f$, 
and $Q_e/Q_u$ from Eq.~(\ref{eq:d}) 
by substituting the maximum value 
of $|S_{21}|$ for $d.$ 
In Fig.~\ref{fig:B1Ql}, for example, 
the sharp decrease of $Q_l$ at $f_0<11$~GHz 
is mostly due to the change in $Q_u$.  
We have also confirmed that the frequency 
dependence of $Q_e$ is consistent with the 
theoretical prediction that $Q_e$ 
is determined mainly by $\omega_0 C_c$. 
The unloaded quality factor $Q_u$ is 
a measure of the internal loss. 
In our resonators, both CPWs and JJs can 
be the source of loss. 
Resonators~A1 and B1 have completely 
different $Q_u$ in the linear regime:  
$Q_u=1.6\times10^4$ and $1.3\times10^3$, 
respectively. 
As the most probable reason, we 
have already mentioned the difference 
in the quality of the dielectric for the CPWs. 
The value of $Q_u=1.6\times10^4$ for Resonators~A1 
is actually still considerably smaller than 
those we obtained for usual CPW resonators 
without JJ fabricated at the same time 
as Resonators~A1. 
Thus, there should be a contribution from JJ as well. 

In the bistable region, it is still possible to 
estimate the quality factors.  
We have seen that Fig.~\ref{fig:A1Pdep}(b) 
suggests a strong $P_{\rm in}$ dependence 
of $Q_u$ in Resonator~A1. 
The $P_{\rm in}$ dependence is likely to be 
due to a finite $R_{\rm qp}$ of JJ. 
The main reason is that the usual CPW resonators 
without JJ looked almost $P_{\rm in}$ independent 
in the relevant power range. 
In general, $Q_u$ of 
CPW resonators is known to increase 
with increasing $P_{\rm in}$ at least 
in the weak-power range, 
where the average energy stored in the   
resonator at $\omega_0$ is comparable 
to $\hbar\omega_0$ or smaller.\cite{OCo08} 
This behavior was also observed in our resonators 
in the linear regime. 
Both in Resonators~A1 and B1, 
$Q_u$ gained several percents when $P_{\rm in}$ was 
increased by 5~dB from $P_c-15$~dBm.  
Figure~\ref{fig:A1Pdep}(b) is explained if 
$R_{\rm qp}$ is a function of $V_J$  
and its value decreases 
with increasing $V_J$, which is quite likely 
(see, for example, Sec.~6.3 of Ref.~\onlinecite{Tin96}).  
Although we are interested in the case that 
the JJ stays in the superconducting branch with 
zero dc voltage, it does not necessarily mean  
that the ac voltage is also zero. The ac voltage 
is given by Eq.~(\ref{eq:Vj}), and when it is nonzero, 
an ac current flows through $R_{\rm qp}$ 
causing a loss. 
This picture is supported by the following 
quantitative consideration: 
according to the simulation in Sec.~\ref{sec:Th}, 
the reduction of $Q_l$ becomes considerable when 
$R_{\rm qp}$ is decreased below $10^3$~$\Omega$ 
in Resonator~A1, whereas  
the dc measurements on the test junctions  
suggest that $R_{\rm qp}$ in Resonator~A1 
can be as low as $3\times10^2$~$\Omega$. 
It is also consistent with the results 
on Resonator~B1. Because B1 has a smaller $J_c$ 
and a much larger superconducting gap, the 
tunnel barrier of the junctions is thicker. 
Thus, $R_{\rm qp}$ must be much larger, 
which explains the fact that 
$P_{\rm in}$ dependence is much smaller. 
The $\Phi$ dependence is also explained 
because $\Phi$ modulates the ratio of 
$R_{\rm qp}/(\omega_0 L_{J0}).$ 
As $\Phi$ approaches $\pm 0.5\Phi_0,$ 
$\pm1.5\Phi_0,$ $\cdots,$ 
$L_{J0}$ increases, but there 
is no obvious mechanism 
that $\Phi$ changes $R_{\rm qp}$. 
We have also confirmed by the simulation 
in Sec.~\ref{sec:Th} that $Q_l$ indeed 
depends on the ratio of 
$R_{\rm qp}/(\omega_0 L_{J0}),$ 
and that the experimental $Q_l$ vs.\ $f_0$ 
in Fig.~\ref{fig:B1Ql} is qualitatively 
reproduced with $R_{\rm qp}$ 
on the order of $10^3 - 10^4$~$\Omega.$ 
In Fig.~\ref{fig:B1Ql}, the predictions based on 
the simulation for different values of 
$R_{\rm qp}$ are also shown by the 
solid and dotted curves.  

From the viewpoint of the qubit readout and the 
qubit coherence time, the dissipation 
due to the loss at $R_{\rm qp}$ 
%
can be unfavorable depending on the circuit 
configuration, 
and worse than that at CPWs when  
$R_{\rm qp}$ is closer to the qubit.  
Fortunately, now we know from the experiment 
in Sec.~\ref{sec:R}, how large $R_{\rm qp}$ 
has to be in order for the loss 
at $R_{\rm qp}$ to be negligible.   
For our resonators in Table~\ref{tab}, 
the ratio of $R_{\rm qp}/(\omega_0 L_{J0})$ 
needs to be larger than $10^3-10^4$ 
according to the simulation in Sec.~\ref{sec:Th}.  
This condition should be easily satisfied 
even with Al JJs by 
employing a sufficiently small 
critical-current density $J_c$.  

\section{Conclusion}
We studied nonlinear superconducting 
resonators with single Josephson 
junctions or dc SQUIDs.  
The bistable region of the resonators 
were experimentally determined, and 
compared with the theory and simulations. 
We found that the variation of the 
unloaded quality factor as a function 
of relevant quantities such as the drive power 
and the external magnetic flux, 
was important for understanding 
the experimental results. 
The unloaded quality factor is a measure of 
the internal loss, and  
the origin of its variation was also discussed.

\section*{Acknowledgment}
The authors thank 
Y. Nakamura for comments, 
K. Matsuba for across-the-board assistance, 
and Y. Kitagawa for fabrication assistance. 
M.\ W.\ thanks R.\ L.\ Kautz for fruitful discussion. 
T.\ Y.\ and J.-S.\ T.\ thank CREST-JST, Japan 
for financial support.


\end{document}